\documentstyle[12pt]{article}
\oddsidemargin--0mm
\begin{document}

\begin{titlepage}

\vskip 4cm
\begin{center}
{\Large\bf  Backreaction of excitations on a  vortex $^{\ast}$} \\

\vskip 1cm

by \\

\vskip 5mm

{\large\bf Henryk Arod\'z}$^{\dag}$ and
{\large\bf Leszek Hadasz}$^{\ddag}$

\vskip 1cm

Jagellonian University, Institute of Physics \\
Reymonta 4, 30--059 Cracow, Poland
\end{center}

\begin{abstract}
Excitations of a vortex are usually considered in a linear approximation
neglecting their backreaction on the vortex. In the present 
paper we investigate  backreaction of Proca type excitations
on a straightlinear vortex in the Abelian Higgs model. We propose exact
Ansatz for fields of the excited vortex. From initial set of six 
nonlinear  field equations we obtain (in a limit of weak excitations) 
two linear wave equations  for the backreaction corrections.
Their approximate solutions are found in the cases of plane wave and 
wave packet type excitations. We find that the excited vortex radiates
vector field and that the Higgs field has a very broad oscillating component.

\end{abstract}

\vspace{\fill}

\noindent
\begin{tabular}{l}
TPJU 10/96 \\
May 1996\\
hep-th/9607116
\end{tabular}

\vskip 2cm
\noindent
\underline{\hspace*{10cm}}

\noindent
$^{\ast}$Work supported in part by grant KBN 2 P302 049 05.

\noindent
$^{\dag}$ E--mail: ufarodz@ztc386a.if.uj.edu.pl.

\noindent
$^{\ddag}$ E--mail: hadasz@thp1.if.uj.edu.pl.

\noindent
\end{titlepage}

\section{ INTRODUCTION} 

String--like vortices play essential role in several branches of physics,
e.g., in condensed matter physics, in physics of hadrons, or in 
field-theoretical cosmology. The vortices have very
interesting, intricate dynamics. Its understanding is crucial for
extracting predictions from many theoretical models. 

  We are interested in a particular aspect of vortex dynamics --
properties of excited vortex. Excitations of vortices have been studied
already for some time, see e.g. [1] for recent results and references to
older papers. Our interest in this aspect of vortex dynamics stems from 
the fact that there is no large energy barrier or exclusion principle for 
production of excited  vortices. Therefore, one may expect that in 
formation or interaction processes excited vortices will occur quite often, 
similarly as excited atoms or molecules are obtained in a recombination 
process. Such expectations are supported by computer simulations of the 
intercommutation of vortices [2], where one observes that parts of vortices 
which are close to the interaction point are locally  excited. 
For concreteness, we consider the Abrikosov-Nielsen-Olesen (ANO) 
vortex [3] in the Abelian  Higgs model with a particular type of axially 
symmetric excitation discussed already in [4], [5]. The excitation can be
described as a bound state of longitudinally polarized vector particles
with the vortex. In the present paper we study the excited vortex from a 
general field-theoretical viewpoint --  applications to vortices in 
superconductors will be presented in a forthcoming paper [6].

The problem we are addressing ourselves to can be described as follows.  
The excitation is usually obtained as a first order correction (in an 
expansion with respect to amplitude of the excitation) to the unexcited ANO 
vortex fields, which are taken as the zeroth order fields. 
Up to the first order approximation the total field configuration is 
just a sum of the fields of unexcited vortex and of the excitation. 
However, because of nonlinearity of the pertinent field equations there 
are higher order corrections to this sum.  Generally, they can be regarded as
giving interaction of the excitation with the unexcited ANO vortex, and also
selfinteraction of the excitation. Such second and higher order effects in the
case of vortices seem  to be unexplored as yet. In the present paper we 
calculate second order corrections in the perturbative expansion  with 
respect to amplitude of the excitation in order to check 
what kind of  physical effects appear.
It turns out that these second order effects are quite interesting.
We find that the excited vortex emits waves of the vector field, and that 
it has larger and oscillating width. All corrections found in the second 
order can be regarded as a backreaction of the excitation on the ANO vortex.
Selfinteraction of the excitation will appear in the third order.

The present paper is based on our earlier work [7] in which we have
obtained approximate analytic formulae for the ANO vortex, as well as for the
excitation profile and frequency ($F^{\rm ANO}$, $\chi^{\rm ANO}$, 
$\alpha(r)$ and $\omega_0$ below). We have also considered there the
backreaction in the case of homogeneously excited vortex, which is a 
particularly simple case because of the lack of dependence on 
position along the vortex. Certain results obtained in [7] 
hold for excitations of finite amplitude, e.g., we have found an 
upper bound on amplitude of the homogeneous excitation. In the present 
paper we try to cope with inhomogeneous excitations. The price for greater 
generality is that we can investigate only excitations which have small 
amplitude  --  our approach is perturbative one.

In order to keep technical details at a reasonable level we  
consider only
elementary vortex with unit topological charge.  The vortex is 
straight-linear,  axially symmetric, and its core lies on the $x^3-$ axis. 
Another simplification we adopt in our
paper is that the parameter $\kappa \equiv \sqrt{2q^2/\lambda} $ is assumed 
to be very small, $\kappa\ll 1/2$. In a static case this corresponds to 
strongly II type superconductors. Actually, in several places we consider the 
limiting case $\kappa \to 0$, but in fact we expect that our conclusion remain
qualitatively valid for all $\kappa < 1/2.$  The point $\kappa =1/2$ is 
distinguished by a change of asymptotic form of the Higgs field [8]  -- a
polynomial approximation used in [7] has not been extended as yet to this and
higher values of $\kappa$.

The plan of our paper is as follows. In the next Section we present
Ansatz and field equations for the axially symmetric excited vortex. 
In Section 3  we describe the perturbative expansion
with respect to the amplitude of excitation. Sections 4 and 5 are 
devoted to detailed analysis of the backreaction in the 
cases of plane wave and  wave packet type excitations. 
In Section 6 we have collected  several  remarks.

\vspace*{1cm}

\section{ THE ANSATZ} 

Our notation is fixed by writing the Lagrangian of the Abelian Higgs model
in the following form, 
\begin{equation} 
{\cal L} =
    \left( D_\mu\phi\right)^{\mbox{$\ast$}}\!\left(D^\mu\phi\right) -
    \frac{\lambda}{4}\left(\phi^{\mbox{$\ast$}}\!\phi -\frac{2m^2}{\lambda}
    \right)^2 -
    \frac{1}{4} F_{\mu\nu}F^{\mu\nu},
\end{equation}
where
$$
    D_\mu\phi = \partial_\mu\phi +iqA_\mu\phi, \hskip 1cm
    F_{\mu\nu} = \partial_\mu A_\nu - \partial_\nu A_\mu.
$$
Signature of the space--time metric is $(+,-,-,-)$ and  $c=1$ as usual.
Masses of the scalar and vector particles are equal to $m_H = \sqrt{2}m$ 
and $m_A = \kappa m_H,$ respectively. In the following we put
$q=1.$

To describe the excited vortex we use the most general axially symmetric
extension of the Abrikosov -- Nielsen -- Olesen (ANO) Ansatz,
\begin{equation}
\phi = \sqrt{\frac{2m^2}{\lambda}}\exp\Big[i\big(\theta + 
           \vartheta(\xi^\alpha,r)\big)\Big]F(\xi^\alpha,r),
\end{equation}
\begin{equation}
A^1 = -  m_H \frac{r^2}{r} \frac{1- \chi(\xi^\alpha,r)}{r},
\hskip 1cm
A^2 =  m_H \frac{r^1}{r} \frac{1- \chi(\xi^\alpha,r)}{r},
\end{equation}
\begin{equation}
A^\beta = m_H A^\beta(\xi^\alpha,r).
\end{equation}
Here we use the dimensionless variables,
$$
\begin{array}{lll}
r^k = m_H x^k, & & r = \sqrt{(r^1)^2 + (r^2)^2}; \\
&& \\
\xi^\alpha = m_H x^\alpha. &  &
\end{array}
$$

The indices $\alpha,\beta$ take values 0, 3, while $i,k = 1,2.$ $\theta$ 
denotes the azimuthal angle in the $(r^1,r^2)$ plane.  The functions 
$F, \chi, A^\beta$ and $\vartheta$ are dimensionless. 
For the unexcited, static ANO vortex
\begin{equation}
\vartheta = 0, \hskip 1cm A^\beta = 0,
\end{equation}
and $F, \chi$ do not depend on $\xi^\alpha.$

The Ansatz (2--4) and Euler--Lagrange equations 
obtained from  Lagrangian (1) lead to the following equations for 
$F,\vartheta,A^\beta$ and $\chi,$
\begin{equation}
-\partial_\beta\partial^\beta F + F'' + \frac{1}{r}F' + 
\left[\big(A_\beta + \partial_\beta\vartheta\big)^2 +
 \frac12\big(1-F^2\big) -
\left(\frac{\chi^2}{r^2} + \vartheta'{}^2\right)\right]F = 0,
\end{equation}
\begin{equation}
F\partial_\beta\partial^\beta\vartheta + 2\partial_\beta F\partial^\beta
\vartheta - 2F'\vartheta' - 
F\left(\vartheta''+ \frac{1}{r}\vartheta'\right) + 2A^\beta\partial_\beta F +
\left(\partial_\beta A^\beta\right)F = 0,
\end{equation}
\begin{equation}
- \partial_\alpha\partial^\alpha A^\beta + \partial^\beta\left(
\partial_\alpha A^\alpha\right) +
\left(A^\beta{}'' + \frac{A^\beta{}'}{r}\right) - \kappa^2F^2A^\beta  = 
   \kappa^2F^2\partial^\beta\vartheta,
\end{equation} 
 \begin{equation}
-\partial_\beta\partial^\beta\chi + \chi'' -\frac{\chi'}{r} = \kappa^2F^2\chi,
\end{equation} 
\begin{equation}
\partial_\beta A^\beta{}' = \kappa^2F^2\vartheta'.
\end{equation}
Here ' stands for $d/dr.$
Equations for the vector field we have split into the $\beta = 0, 3$
and $i = 1, 2$ components.

Equation (7) has the form of wave equation for $\vartheta$.
It turns out that it may be omitted  because it follows from Eqs. (8), (10).
To check this, act with $\partial_\beta$ on  both sides of Eq. (8) to obtain
$$
\partial_\beta A^\beta{}'' = -\frac{1}{r}\partial_\beta A^\beta{}' + 
  2\kappa^2F\partial_\beta F\left(\partial^\beta\vartheta + A^\beta\right) +
  \kappa^2F^2\left(\partial_\beta\partial^\beta\vartheta +
   \partial_\beta A^\beta\right).
$$
Next, eliminate $\partial_\beta A^\beta{}'$ from this formula using the 
relation (10) and divide the resulting formula by $\kappa^2 F$. 
The result coincides with Eq. (7).

Thus, in addition to Eqs. (6), (9) which are 
nontrivial also for the unexcited ANO vortex (cf. formulae (5)),
we have two nonlinear wave equations more, namely the two components of 
Eq. (8).  The phase field $\vartheta$  is determined from  Eq. (10):
\begin{equation}
 \vartheta(\xi,r) = -  \kappa^{-2} \int^{\infty}_{r} ds \; F^{-2}(\xi,s)
\partial_{\beta}A'^{\beta}(\xi,s).   
\end{equation}
This formula implies a gauge fixing for a residual gauge freedom present in 
the Ansatz (2-4) and consisting of gauge transformations with gauge
function $\delta\vartheta$ dependent only on $\xi^{\alpha}$. Using this 
freedom we may adjust $\vartheta$ so that it vanishes for $ r \to \infty$ -- 
this is the gauge implied by formula (11). Notice that in spite of U(1) gauge
invariance of Lagrangian (1) the phase $\vartheta(\xi^\alpha,r)$
cannot be gauged away within the Ansatz (2-4) because the corresponding 
gauge transformation would introduce terms $\delta A^i \sim r^i/r$ which are 
not compatible with formulae (3) for $A^i.$ In an alternative 
form of the Ansatz we could have put $\vartheta = 0$ and introduced the 
terms $\sim r^i/r$ in formulae (3).

Apart from the wave equations, we also require that the functions 
$F$, $\chi$, $A^{\beta}$  obey a number of boundary conditions.  
For $F$  and $\chi$  at $r=0$  we take the standard ANO vortex conditions
\begin{equation}
F(0) = 0 , \hskip 1cm \chi(0) = 1,
\end{equation}
which follow from the requirement that the fields $\phi$ and $A^{i}$ are
regular at $r=0$ -- we are not interested here in singular solutions.
Boundary conditions at $r \to \infty$ are less obvious. First, in 
our perturbative approach the pair of equations (6, 9) is replaced by 
infinite set of equations, a pair in each order. It turns out that it is not 
always possible to impose the conditions that the perturbative contributions 
to $F$ and $\chi$ exponentially vanish for $r \to \infty$. The reason is that
most of the pertubatively obtained equations  have radiation type 
solutions which  vanish very slowly or do not vanish at all for large 
$r$. In this case we adopt the Helmholtz condition which states that for 
$ r\to \infty$ only outgoing radiation waves are present, [9]. 

As for the functions $A^\beta$, we shall require that they are regular for 
all $r\geq 0 $ and that they vanish for $r \to \infty$. These conditions
correspond to the physical picture that the excitation is localised on the 
vortex, i.e. that it can be regarded as a bound state of the $\beta = 0, 3$ 
components of the vector  field $A_{\mu}$  with the vortex.  

\vspace*{1cm}

\section{ THE PERTURBATIVE EXPANSION} 

The set of nonlinear wave equations (6), (8), (9) is very complicated. 
In the following part of our paper we attempt to construct its perturbative
solution relevant for the excited vortex in the case when the excitation, 
represented by the $A^\beta$ fields, has a small amplitude which will be 
regarded as expansion parameter. The ensuing picture
is as follows. In the leading order $(\varepsilon^0)$ we have the 
unperturbed ANO vortex.
In the order $\varepsilon^1$ waves of the $A^\beta$ field travelling along
the vortex appear. They can be described as a 2--dimensional free Proca 
field living on the vortex. In the next order $(\varepsilon^2)$ backreaction
of the Proca wave on the vortex appears. It introduces two new features: 
perturbation of radius of the vortex and  radiation from the vortex.

The perturbative expansion has the form
\begin{equation}
 A_\alpha = \varepsilon A^{(1)}_\alpha + {\cal O}(\varepsilon^3),
\end{equation}
\begin{equation}
 F = F^{\rm ANO} +\varepsilon^2 F^{(2)} + {\cal O}(\varepsilon^4),
 \hskip 5mm
 \chi = \chi^{\rm ANO} + \varepsilon^2\chi^{(2)} + {\cal O}(\varepsilon^4), 
\end{equation}
where $F^{\rm ANO},$ $\chi^{\rm ANO}$ give the well-known static, unexcited 
ANO vortex. 
Approximate formulae for $F^{\rm ANO},$ $\chi^{\rm ANO}$ are given below.
$\varepsilon$ is just an auxilliary book--keeping parameter --- it is put 
to 1 at the end of calculations. The power expansion in 
$\varepsilon$ is what we mean by the expansion in  amplitude of the 
excitation $A^{(1)}_\alpha.$ The powers of $\varepsilon$ in formulae (13, 14)
are suggested by the form of Eqs. (6), (8), (9).

Let us first eliminate the phase $\vartheta$. In the leading order, 
Eq. (10) admits as the solution
\begin{equation}
   \vartheta^{(1)} = 0,
\end{equation}    
with
\begin{equation}
  \partial^\beta A^{(1)}_\beta = 0.
\end{equation} 
This solution  is compatible with Eq. (8) because 
$\partial_{\beta} F^{\rm ANO} = 0. $ 

Now, with (15) taken into account, Eq. (8) in the order 
$\varepsilon^1$ decouples from the other equations,
\begin{equation}
-\partial_\alpha\partial^\alpha A^{(1)}_\beta +A^{(1)}_\beta{}'' + 
\frac{1}{r}A^{(1)}_\beta{}' - \kappa^2(F^{\rm ANO})^2A^{(1)}_\beta =0.
\end{equation} 
This equation has the form  of 4-dimensional wave equation and as such it 
has plenty of solutions. We are interested in bound state type solutions, 
i.e. the ones which exponentially vanish for large $r$. They can be found 
with the  help of separation of variables $\xi^\beta$ from $r,$
\begin{equation}
 A^{(1)}_\beta = W_\beta(\xi)\alpha(r),
\end{equation} 
where $W_\beta(\xi)$ obey the Lorentz condition
\begin{equation}
 \partial_\beta W^\beta = 0,
\end{equation}
and the 2--dimensional Proca equation
\begin{equation}
 \partial_\alpha\partial^\alpha W_\beta + \omega_0^2W_\beta = 0
\end{equation}
with the mass $\omega_0$  given below.

It follows from (17--20) that the profile function 
of the excitation, i.e. $\alpha(r),$  obeys the equation
\begin{equation}
\omega_0^2\alpha(r) + \alpha''(r) + \frac{\alpha'(r)}{r} -
 \kappa^2(F^{\rm ANO})^2\alpha(r) = 0.
\end{equation}
This equation has been  considered in [5], [7]. Its vanishing for 
$r \to \infty$ solution is quoted below. Without any loss of generality we 
may normalize $\alpha(r)$. The solution given below is normalized by 
the condition
\[ \alpha(0)=1.  \]

Proca equation (20) possesses infinitely many solutions. Among them there
are travelling plane waves and wave packets -- 
they give rise to the inhomogeneous
excitations considered in the present paper. The "breathing" homogeneous
vortex considered in [7] is obtained for $W^0 = 0, W^3 = W^3(x^0),$ 
which is a non--propagating,
periodic in time and $x^3$-independent Proca field standing on the vortex.
Condition (19) implies that there exists a potential $\psi$ for the 
$W^{\beta}$ field,
\[ W_{\beta} = \epsilon_{\beta \alpha} \partial^{\alpha} \psi, \]
where $ \epsilon_{\beta \alpha} = - \epsilon_{\alpha \beta}$ and 
$\epsilon_{03}=+1$. Proca equation (20) will be satisfied if 
the potential obeys Klein-Gordon equation
\begin{equation}
\partial_{\gamma}\partial^{\gamma} \psi + \omega_{0}^{2} \psi = 0.
\end{equation}
In the following Sections we shall consider the plane wave solution of 
equation (22), namely
\begin{equation}
\psi = N \sin(E(q)\xi^0 - q \xi^3 +\delta_0),
\end{equation}
where $E(q)=\sqrt{\omega_0^2 +q^2},$ $N$ is a constant, and $\delta_0$ 
is a constant phase which we
shall put to zero. 

Another  solution of Eq. (22), also  considered in the
following Sections, is an approximate solution giving a standing wave packet,
\begin{equation}
\psi \approx N \exp(-\xi^2_3/4\Lambda^2) \sin(\xi^0 \omega_0),
\end{equation}
where $\Lambda$ and $N$ are constants. Average momentum of this wave packet 
is equal to zero. We  also assume that $\omega_0 \Lambda 
\gg 1$. Then, the wave packet has momentum cutoff at small momentum $k_3 \sim 
\Lambda^{-1} \ll 1$. In this case we may neglect (for a finite time $0 \leq
\xi^0 < \Lambda^2$) dispersion of the wave packet, 
which is of course present in exact wave  packet solutions of Eq. (22).
Such exact wave packet solutions are easy to obtain, nevertheless
we prefer the approximate solution because it has 
significantly simpler form.

Now, let us turn to the remaining two equations (6) and (9). In
the order $\varepsilon^0$ they are solved by the unexcited vortex 
fields $F^{\rm ANO},  \chi^{\rm ANO}$. In the order 
$\varepsilon^2$ they give the following equations for $\chi^{(2)},F^{(2)}:$
\begin{eqnarray}
- \partial_\beta\partial^\beta F^{(2)} + F^{(2)}{}'' + \frac{1}{r}F^{(2)}{}'
 +   \frac12 F^{(2)}\left[1-3(F^{\rm ANO})^2 - 
  \frac{2}{r^2}(\chi^{\rm ANO})^2\right] =
  \hspace*{3cm} &&  \\
  \hspace*{3.0cm}
  \frac{2}{r^2}\chi^{\rm ANO}F^{\rm ANO}\chi^{(2)}  
        -F^{\rm ANO}W_\beta W^\beta\alpha^2(r), & &   \nonumber 
\end{eqnarray}
\begin{equation}
  -\partial_\beta\partial^\beta\chi^{(2)} + \chi^{(2)}{}'' -
   \frac{1}{r}\chi^{(2)}{}'
  -\kappa^2(F^{\rm ANO})^2\chi^{(2)} =
  2\kappa^2F^{\rm ANO}\chi^{\rm ANO}F^{(2)}. 
  \end{equation}
The boundary conditions for $F^{(2)},\chi^{(2)}$ at $r=0$ follow from 
conditions (12) and from the fact that $F^{\rm ANO},\chi^{\rm ANO}$ already
obey these conditions. Therefore, 
\begin{equation}
F^{(2)}(r=0,\xi^\alpha) = 0, \hskip 1cm
\chi^{(2)}(r=0,\xi^\alpha) = 0.
\end{equation}
We shall see that we may also require that
\begin{equation}
 F^{(2)}(r,\xi^{\alpha}) \to 0 \;\;\;  \mbox{for} \;\;\; r \to \infty. 
\end{equation}
As for  $\chi^{(2)}$, we find that Eq. (26) has radiation type solutions, so 
we adopt the outgoing  radiation condition.

The last term on the r.h.s of Eq. (25)  is a source term for $F^{(2)}$. 
Due to its presence  $F^{(2)}, \chi^{(2)}$  do not vanish. Hence, the 
backreaction of the excitation on the vortex is always present (one can 
easily prove that $W_{\beta}W^{\beta} \neq 0$ if $W_{\beta} \neq 0).$

From a mathematical point of view, Eqs. (25), (26) form a set of
4--dimensional, linear wave equations with the coefficients 
given by the ANO functions $F^{\rm ANO},\chi^{\rm ANO}.$ Solutions of such
equations can be investigated with a help of numerical methods. In the 
present  paper we take another route. We will present approximate
analytical solutions which we can obtain in the limiting case 
$\kappa \to 0$, i.e. when $0 < \kappa \ll 1/2$ (in practice we mean $\kappa$ 
not greater than 0.1). This limit is distinguished by two significant 
simplifications. 

 First, because  $\kappa < 1/2 $ we may use relatively simple approximate
formulae for the functions $F^{\rm ANO}, \chi^{\rm ANO}, \alpha(r)$ and for 
the frequency $\omega_0$ obtained in [7]. In the limiting case 
$\kappa \ll 1/2 $ we obtain from formulae given in [7] that for 
$r\leq r_0 = 4/\sqrt{3}$
\begin{eqnarray}
\chi^{\rm ANO} & \simeq & 1, \\
&& \nonumber \\
F^{\rm ANO} & \simeq & \frac{3}{2r_0}r - \frac{1}{2r_0^3}r^3, \\
&& \nonumber \\
\alpha(r) & \simeq & 1,
\end{eqnarray}
while for $r\geq r_0$
\begin{eqnarray}
\chi^{\rm ANO} & \simeq & \kappa rK_1(\kappa r), \\
&& \nonumber \\
F^{\rm ANO} & \simeq & 1, \\
&& \nonumber \\
\alpha(r) & \simeq & c_1K_0(k_0r),
\end{eqnarray}
where
\begin{equation}
c_1 \simeq \frac{23}{3} \kappa^2,
\end{equation}
and
\begin{equation}
k_0 \simeq 0.85 \; \exp \left\{-\frac{3}{23}\frac{1}{\kappa^2}\right\}.
\end{equation}
The frequency $\omega_0$ is given by the formula
\[ \omega_0 = \sqrt{\kappa^2 - k_0^2} .  \]
$K_0(z)$ and $K_1(z)$ are the modified Hankel functions, [10]. The 
corresponding functions in the two regions smoothly match each 
other at $r=r_0$. On the r.h.s. of formulae (29-36) we have neglected terms 
of the order $\kappa^2 \ln\kappa$ or smaller. More accurate formulae for the 
ANO vortex and the excitation can be found in [7]. 
Notice that  $k_0$ is exceedingly small, and therefore the profile function
$\alpha(r)$ is very broad, much broader than $\chi^{\rm ANO}$ and 
$F^{\rm ANO}$ which have the width $\sim 1/\kappa, \;\; \sim r_0$, 
respectively.

The second simplification occuring for $\kappa \ll 1/2$ is  that
the first term on the r.h.s. of Eq. (25) can be neglected. To see this, first 
notice that this term is finite at $r=0$ 
because $F^{\rm ANO}$ and  $\chi^{(2)}$ vanish at that point. 
Now, the r.h.s. of Eq. (26) is of the order 
$\kappa^2$, therefore $\chi^{(2)}$ determined from that equation is expected
to vanish for $\kappa \to 0$. Therefore, the discussed term for the 
very small $\kappa$ will be negligibly small in comparison with the 
other term on the r.h.s. of Eq. (25) which stays finite in the limit 
$\kappa \to 0$.   Thus, instead of Eq. (25) we may consider the equation
\begin{eqnarray}
- \partial_\beta\partial^\beta F^{(2)} + F^{(2)}{}'' + \frac{1}{r}F^{(2)}{}' +
\frac12 F^{(2)}\left[1-3(F^{\rm ANO})^2 - 
\frac{2}{r^2}(\chi^{\rm ANO})^2\right] =
\hspace*{3cm} &&  \\
  \hspace*{2cm} -F^{\rm ANO}W_\beta W^\beta\alpha^2(r). && \nonumber 
\end{eqnarray}
Due to this latter simplification the set of intercoupled equations (25),
(26) is replaced by equations (37) and (26) which can be solved one after
the other.
$F^{(2)}$ determined from Eq. (37) gives the source term  in Eq. (26).   
In the following two Sections we shall present solutions of these 
equations in the particular cases of the excitation field $W^{\beta}$
given by the plane wave (23) and the wave packet (24).

\section{ BACKREACTION ON THE HIGGS FIELD }

In this Section we shall analyse  the backreaction  of the excitations  on 
the Higgs field $F$. Our starting point is the equation (37).

It is convenient to pass to Fourier transforms with respect to $\xi^{\alpha}$,
\begin{equation}
F^{(2)}(r,\xi) = \frac{1}{2\pi} \int d\omega dk \; 
\exp(-i\omega \xi^0 + ik\xi^3) f(r,k,\omega),
\end{equation}
\begin{equation}
\chi^{(2)}(r,\xi) = \frac{1}{2\pi} \int d\omega dk \exp(-i\omega \xi^0 + 
ik\xi^3) h(r,k,\omega),
\end{equation}
and
\begin{equation}
w(k,\omega) = \frac{1}{2\pi} \int d\xi^0 d\xi^3 \;
 \exp(i\omega \xi^0 - ik\xi^3) W_{\beta}(\xi) W^{\beta}(\xi).
\end{equation}
The Fourier transform of Eq.(37) has the following form
\begin{eqnarray}
f'' + \frac{1}{r} f' - \left[ \omega^2 - k^2 - \frac{1}{2} 
+ \frac{3}{2}(F^{\rm ANO}(r))^2 + \frac{1}{r^2} (\chi^{\rm ANO}(r))^2 
\right] f = \hspace*{2cm}  && \nonumber \\
&& \\
 \hspace*{2cm}  - F^{\rm ANO}(r) \alpha^2(r) w(k,\omega). &&
\nonumber
\end{eqnarray}
For the plane wave (23) with $\delta_0 = 0$
\[  W_{\beta}(\xi) W^{\beta}(\xi) = - N^2 \omega^2_0 
\cos^2(E(q)\xi^0 - q \xi^3),  \]
and  
\begin{equation}
w(k, \omega) = - \pi \omega_0^2 N^2 \left[ \delta(\omega) \delta(k) +
\frac{1}{2} \delta(\omega - 2 E(q)) \delta(k-2 q) + 
\frac{1}{2} \delta(\omega + 2 E(q)) \delta(k+2 q) \right].
\end{equation}
In the case of the wave packet (24)
\begin{equation}
W^{\alpha}W_{\alpha} = w_1(\xi_3) + w_2(\xi_3) \cos(2 \omega_0 \xi^0),
\end{equation}
where 
\begin{equation} 
w_1(\xi_3) = \frac{1}{2} N^2 (\frac{\xi_3^2}{4\Lambda^4} - \omega_0^2)
\exp(-\frac{\xi_3^2}{2\Lambda^2}),
\end{equation}
\begin{equation}
w_2(\xi_3) =- \frac{1}{2} N^2 (\frac{\xi_3^2}{4\Lambda^4} + \omega_0^2)
\exp(-\frac{\xi_3^2}{2\Lambda^2}). 
\end{equation}
Formula (40) gives in this case
\begin{equation}
w(k, \omega) = \sqrt{2 \pi} \left[ \delta(\omega) \tilde{w_1}(k) +
\frac{1}{2} \delta(\omega - 2 \omega_0) \tilde{w_2}(k) + 
\frac{1}{2} \delta(\omega + 2 \omega_0) \tilde{w_2}(k) \right],
\end{equation}
where $\tilde{w}_{1,2}$ denote Fourier transforms of 
$w_{1,2}(\xi_3)$. We shall not need their explicit form.
  
It is clear from Eq.(41) that  $f$ has the form analoguous to (42), (46), 
respectively, 
\begin{eqnarray}
f =   \pi \omega_0^2 N^2 \left[ f_0(r) \delta(\omega) 
\delta(k) +  \frac{1}{2}    f_+(r)
 \delta(\omega - 2 E(q)) \delta(k-2 q) \right.  & &   \\ 
\left.  + \frac{1}{2}  f_-(r)  \delta(\omega +
 2 E(q)) \delta(k+2 q) \right], & &
 \nonumber
\end{eqnarray}
or 
\begin{equation}
f =  - \sqrt{2 \pi} \left[ f_0(r) \delta(\omega) \tilde{w_1}(k) +
\frac{1}{2} f_+(r) \delta(\omega - 2 \omega_0) \tilde{w_2}(k) + 
\frac{1}{2}  f_-(r) \delta(\omega + 2 \omega_0) \tilde{w_2}(k) \right].
\end{equation}
The negative and positive frequency components  are related by 
complex conjugation,
\[   f_- =   f_+^{\ast} \]
and $f_0$ is real valued. 

Next, we observe that  the expression in square bracket on the l.h.s. of 
Eq. (41) can be simplified a little bit.  Namely, the term $\omega^2 - k^2$ 
is equal to either 0 or $4\omega_0^2$ in both the plane wave and wave packet 
cases, and for small $\kappa$ it is negligibly small in comparison with 
the other terms in the square bracket -- a plot of the function  $- 1/2 
+ 3/2(F^{\rm ANO}(r))^2 + (\chi^{\rm ANO}(r))^2/ r^2 $ shows
that its minimal value is approximately equal to 0.945. Therefore, instead 
of Eq. (41) we may consider the following universal equation, common for all 
frequency components and for the plane wave and wave packet cases,
\begin{equation}
f_u'' + \frac{1}{r} f_u' - \left[ - \frac{1}{2} 
+ \frac{3}{2}(F^{\rm ANO}(r))^2 + \frac{1}{r^2} (\chi^{\rm ANO}(r))^2 
\right] f_u =  F^{\rm ANO}(r) \alpha^2(r),
\end{equation}
where 
\begin{equation}
f_u = f_0 \approx f_{\pm}.
\end{equation}

In order to see the backreaction on the Higgs field  we have to solve Eq. (49)
for $f_u$. To this end we shall use the polynomial approximation which has 
proved useful also in our earlier work [7].  The idea is to find approximate
solutions of Eq. (49) separately in the regions $r \leq r_0$ and $r \geq r_0$,
and to glue them at $r = r_0$. 
 
In the "outer" region, i.e. for $r \geq r_0 = 4/\sqrt{3}$, we use
formulae (32-36). Thus,  $F^{\rm ANO}(r) \simeq 1$, and the functions 
$ (\chi^{\rm ANO})^2/r^2$ and $\alpha(r)^2$ are smooth and  very slowly 
changing with $r$. Their derivatives with respect to $r$ are of the order 
$\kappa$ and $k_0$, respectively, hence they are close to zero. Therefore,  
the following function is a reasonable approximation to the exact 
solution of Eq. (49) in the outer region
\begin{equation}
f_u = \tilde{f}_u(r) \equiv - \alpha(r)^2 \left( 1 + 
\frac{(\chi^{\rm ANO}(r))^2}{r^2} \right)^{-1} + f_0 K_0(r),
\end{equation}
where the last term on the r.h.s. is a general, vanishing for $r \to \infty,$
solution of the homogeneous counterpart of Eq. (49). $f_0$ is a constant 
to be determined from  matching conditions at $r = r_0$.

On the other hand, for $r\leq r_0$  the ANO functions   
are given by polynomials (29-31), and we seek a polynomial approximation also 
for $f_u$, 
\begin{equation}
f_u = \underline{f}_u  \equiv f_1 r + f_3 r^3 + f_5 r^5. 
\end{equation} 
We take the fifth order polynomial for $f_u$ because formula (30) gives
$F^{\rm ANO}$  up to the $r^3$ term. Inclusion of the next 
term $( \sim r^7)$ would make sense if we knew $F^{\rm ANO}$ up to
 $\sim r^5$ term, as it is clear from comparison of the both sides of 
Eq. (49).
Equation (49) gives the following  relations
\[  f_3 = \frac{3}{16r_0} - \frac{1}{16} f_1,  \] 
\[ f_5 = - \frac{1}{48 r_0^3} + \frac{9}{64 r_0^2} f_1 - \frac{1}{48}f_3, \]
where $f_1$ remains undetermined.

At $r=r_0$ we impose matching conditions. They have the following form
\[  \underline{f}_u (r_0) = \tilde{f}_u (r_0), \;\; \; 
 \underline{f}_u'(r_0) = \tilde{f}_u'(r_0). \]
These conditions give a set of linear, algebraic equations for the constants 
$f_0, f_1$ -- the solution is 
\begin{equation}
 f_0 \approx 5.19, \;\;\;  f_1 \approx  - 0.36.  
\end{equation}

From formulae (38), (47), (48), (50) we obtain the following final expression
for  the backreaction on the scalar  field
\begin{equation}
F^{(2)} \approx - W^{\alpha}W_{\alpha}(\xi) f_u(r),
\end{equation}
where $f_u(r)$ is given by (51-53).
Formula (54) covers the plane wave and wave packet cases. We see that 
$F^{(2)}$ introduces modulated along the vortex and oscillating in time
very broad  component in the scalar field of the vortex.
Its range is $\sim (k_0 m_H)^{-1}$.

\section{ BACKREACTION ON THE VECTOR FIELD }

Now let us investigate the backreaction on the vector field. Fourier 
transform of equation (26) has the following form
\begin{equation}
h'' - \frac{1}{r} h' + \left[\omega^2 -k^2  
- \kappa^2 (F^{\rm ANO}(r))^2  \right] h = 2 \kappa^2
 F^{\rm ANO}(r) \chi^{\rm ANO}(r) f(r,k,\omega).
\end{equation}
We shall consider the plane wave and the wave packet cases separately.

\subsection{ The plane wave case}
In analogy to formula (47) we write
\begin{eqnarray}
h = 2 \kappa^2  \pi \omega_0^2 N^2 \left[ h_0(r) \delta(\omega) 
\delta(k) +  \frac{1}{2}    h_+(r)
 \delta(\omega - 2 E(q)) \delta(k-2 q)  \right.   & & \hspace*{3cm} \\
     \left.  + \frac{1}{2}  h_-(r)
 \delta(\omega + 2 E(q)) \delta(k+2 q) \right]. & & \nonumber 
\end{eqnarray}

Equation (55) is equivalent to the following set of equations (one equation
for each frequency component),
\begin{equation}
h_a'' - \frac{h_a'}{r} + \left[\Omega_a^2   
- \kappa^2 (F^{\rm ANO}(r))^2  \right] h_a = 
 F^{\rm ANO}(r) \chi^{\rm ANO}(r) f_u(r),
\end{equation}
where the index $a$ takes values $0, +,-,$ and $\Omega_0 =0, \;\;
\Omega_{\pm} = \pm 2 \omega_0$. Very important difference with the case of 
Higgs backreaction $f$ is that
in these equations the $4\omega_0^2 $ term  is of the same order of 
magnitude as the  $\kappa^2 (F^{\rm ANO})^2$ term (recall that
$\omega_0^2 \approx \kappa^2$).

There are two essentially different cases:
$\Omega_0^2 =0$ and $\Omega_{\pm}^2=4\omega_0^2$. In the first case, in 
the outer region ($r \geq r_0$) Eq. (57) has the form 
\begin{equation}
\tilde{h}_0'' - \frac{1}{r} \tilde{h}_0' - \kappa^2 \tilde{h}_0 =
\chi^{\rm ANO}(r) \tilde{f}_u(r),
\end{equation} 
where $\chi^{\rm ANO}$ is given by formula (32). Applying standard methods,
see e.g. [11], we find the following exact, vanishing at $r \to \infty,$  
solution 
\begin{equation}
\tilde{h}_0(r) = r g_1(r) I_1(\kappa r) +  r g_2(r)K_1(\kappa r) + 
b_0 \kappa r K_1(\kappa r),
\end{equation}
where 
\begin{equation}
g_1(r) = - \int^{\infty}_{r} dr'\; \chi^{\rm ANO}(r') K_1(\kappa r')
 \tilde{f}_u(r'), \;\; 
 g_2(r) = - \int^{r}_{r_0} dr'\; \chi^{\rm ANO}(r') I_1(\kappa r')
 \tilde{f}_u(r').
\end{equation}
Here $I_1$ is the modified Bessel function [10]. The constant $b_0$ will be
determined from matching conditions at $r = r_0$.
We see that for  very large $r$, i.e. for $r \gg k_0^{-1}$,
\[ \tilde{h}_0(r)  \sim  \sqrt{r}  \exp(- \kappa r). \]
This component of the backreaction on the vector field is relatively
uninteresting. It gives a small, time- and $\xi^3$-independent, correction
to the $\chi^{\rm ANO}(r)$ field of the same range as 
$\chi^{\rm ANO}$ (the penetration depth 
$\sim \kappa^{-1} m_H^{-1}$). 

Much more interesting are the other components, i.e. the ones with 
$\Omega_{\pm} = \pm 2  \omega_0$.  In this case, Eq. (57) reduces in the outer
region to
\begin{equation}
\tilde{h}_{\pm}'' - \frac{1}{r} \tilde{h}_{\pm}'+ (4 \omega_0^2 - \kappa^2)
\tilde{h}_{\pm} =
\chi^{\rm ANO}(r) \tilde{f}_u(r),
\end{equation}
where again $\chi^{\rm ANO}$ is given by formula (32). 
Applying the standard methods [11], we obtain the solution
\begin{eqnarray}
\tilde{h}_{\pm} = r h_1(r) H^{(1)}_1(\sqrt{4\omega_0^2 - \kappa^2}\; r) +
r h_2(r) H^{(2)}_1(\sqrt{4\omega_0^2 - \kappa^2}\; r)  & &  \nonumber  \\
& & \; \\ +
b_{\pm}  \sqrt{4\omega_0^2 - \kappa^2}\; r 
H^{(1,2)}_1(\sqrt{4\omega_0^2 - \kappa^2}\; r), & & \nonumber 
\end{eqnarray}
where $H_1^{(1)}, H_1^{(2)}=(H_1^{(1)})^* $ are Hankel functions, and
\begin{equation}
h_1(r) = \frac{i\pi}{4} \int^{\infty}_{r} dr'\;
H^{(2)}_1(\sqrt{4\omega_0^2 - \kappa^2} \; r') \chi^{\rm ANO}(r') 
\tilde{f}_u(r'),
\end{equation}
\[ h_2(r) = (h_1(r))^{*},  \]
In order to keep $\chi$ real-valued we have to assume that  
$b_{+} = (b_{-})^{*}$.
In the last term on the r.h.s. of formula (62) one should take $H^{(1)}_1 
(H^{(2)}_1)$ in the positive (negative) frequency component in order to 
satisfy the outgoing radiation condition. 

The functions $h_{1,2}$ 
exponentially vanish for large $r$. On the other hand, for large $z$ [10]
\begin{equation}
H^{(1)}(z) \cong \sqrt{\frac{2}{\pi z}} \exp(i(z-3\pi/4)), 
\end{equation}
hence the last term on the r.h.s. of formula (62) grows like $r^{1/2}$. 
It gives the radiation of the vector field from the vortex.

In the small $r$ region ($r \leq r_0$), for the all frequencies, we 
shall be satisfied with 
approximate polynomial solutions in the form 
\begin{equation}
\underline{h}_{0,+} = \frac{1}{2} d_{0,+} r^2 + \frac{3f_1}{16r_0} r^4 +
\frac{1}{16r_0} (f_3 -\frac{f_1}{16}) r^6,
\end{equation} 
\[\underline{h}_- = (\underline{h}_+)^*.\]
They have been obtained by inserting
the formulae (29, 30, 52) in Eq. (57), and by neglecting all terms 
$\sim \kappa^2$  on the l.h.s. of it (let us remind that we consider the
case $\kappa \ll 1/2 $). 
Again, we have taken a polynomial of the maximal order 
compatible with the fact 
that for small $r$ the r.h.s. of Eq.(57) is known up to terms of the order 
$r^4$.

The constants $b_0, b_{\pm}, d_0, d_{\pm}$ 
are determined from  matching conditions at $r = r_0$:
\[ \underline{h}_{0,\pm}(r_0) = \tilde{h}_{0,\pm}(r_0), \;\;\; 
\underline{h}_{0,\pm}'(r_0) = \tilde{h}_{0,\pm}'(r_0). \]
It turns out that the first two terms on the r.h.s. of formula (62)
computed for $r = r_0$ are of the order $\kappa \ln(\kappa)$ 
and therefore they  can be neglected for small $\kappa$. 
The argument of the Hankel function $H^{(1)}_1$ is very small for $r=r_0$, 
so we may use the following approximate formula [10]
\[ H^{(1)}_1 (z) \approx - i 0.64 / z. \]  
Similarly, \[ \kappa r_0 K_1(\kappa r_0) \approx 1.\]
The resulting from the matching conditions sets of linear algebraic
equations, simplified as described above, give 
\begin{equation}
d_0 \approx \kappa g_1(r_0) - \frac{1}{2} f_1 r_0 - \frac{3}{8},\;\;\;
b_0 \approx - \frac{5}{9} f_1r_0 - \frac{2}{3}, 
\end{equation}
\begin{equation}
d_{\pm} \approx - \frac{8}{3}f_1 r_0^{-1} - 2 r_0^{-2},
\end{equation}
and
\begin{equation}
 b_{\pm} \approx \pm 1.56 i b_0 \mp \frac{1}{2} i \pi \kappa^{-1} 
 (4\omega_0^2 - \kappa^2)^{-1/2} C(\kappa),
\end{equation}
where $C(\kappa)$ is a constant given by the following formula
\[ C(\kappa) = \int^{\infty}_{\kappa r_0} ds \;
 J_1(\sqrt{3-4k_0^2\kappa^{-2}}\;s) s K_1(s) \tilde{f}_u(s/\kappa). \]
In formula (68) only  the second term on the r.h.s. is important
because $\kappa^{-1}$ is very large.
The constants $r_0$ and $f_1$ we know from Sections 3 and 4. 
This completes the determination of the backreaction on the vector field.

The most interesting implication of these formulae is that the backreaction
includes radiation from the excited vortex. The $b_{\pm} $ components of
$\tilde{h}_{\pm}$ inserted in formulae (56), (39) give the
following contribution to $\chi^{(2)}$ for $ r \gg r_0$ 
\begin{equation}
\delta \tilde{\chi}^{(2)} \approx - \kappa  \omega_0^2  
(4 \omega_0^2 - \kappa^2)^{- 1/4} N^2 \sqrt{\pi/2} C(\kappa)
 \sqrt{r} \cos( - 2 E(q)\xi^0 + 2 q \xi^3
+ \sqrt{4\omega_0^2 - \kappa^2}\; r - \pi/4),
\end{equation}
where $E(q) = \sqrt{\omega_0^2 + q^2}$.  This  radiation wave has a  conical
shape -- at fixed time the surface of constant phase, determined by the
following condition
\[ 2 q \xi^3 + \sqrt{4\omega_0^2 - \kappa^2}\; r = \pi/4 +
2 E(q)\xi^0 + const, \]
is a cone with its top point lying on the $\xi^3$ axis.
For $q=0$ the cone degenerates to a cylinder of constant $r$.
Opening angle $\gamma_0$ of the cone, given by the formula
\[ \tan \gamma_0 = \frac{2 q}{\sqrt{4\omega_0^2 - \kappa^2}}, \]
increases with the wave number $q$ of the Proca wave on the vortex.
The wave four-vector corresponding to the radiation wave (69) 
has the following
components 
\[ (k^{\mu}) = \left( 2 E(q), \vec{k} \right),  \]
where 
\[ \vec{k} =  \sqrt{4\omega_0^2 - \kappa^2}  \left( r^1/r,\; r^2/r,\; 
\tan \gamma_0  \right).  \]
Its 4-dimensional invariant length is equal to $\kappa^2,$ in accordance 
with the fact that the vector particle has the mass equal to $\kappa$ 
in the $m_H$ units.

The energy flux related to that conical wave is given by the Poynting vector
\[ \vec{S} = \delta\vec{E} \times \delta\vec{B}, \]
where $ \delta\vec{E}, \delta\vec{B}$ are the contributions to electric and 
magnetic fields corresponding to $\delta \chi^{(2)}$. Simple calculation gives
\begin{equation}
\vec{S} = \pi C(\kappa)^2 \kappa^2 \omega_0^4 (4\omega_0^2 - \kappa^2)^{- 1/2}
 N^4 m_{H}^4 E(q) \frac{1}{r}  \sin^2(k^{\mu}x_{\mu} m_H) \vec{k}  .
\end{equation}

\subsection{ The wave packet case}
Calculations of backreaction of the Proca wave packet on the vector
field $\chi$ are carried out in the same steps as in the plane wave case.
Instead of formula (56) we have now
\begin{eqnarray}
h = - 2 (2\pi)^{1/2} \kappa^2 \left[ \tilde{w}_1(k) \delta(\omega) h_0(r)
+ \frac{1}{2} \tilde{w}_2(k) \delta(\omega - 2 \omega_0) h_+(r) \right. & & \\
\left.+ \frac{1}{2} \tilde{w}_2(k) 
\delta(\omega + 2 \omega_0) h_-(r) \right]. & & 
\nonumber
\end{eqnarray}
Equations for the frequency components $h_0,\; h_{\pm}$ coincide with Eq. 
(57),
where $\Omega^2_a$ is replaced by $- k^2$ or $4 \omega^2_0 - k^2$. However, 
taking into account the assumption that the wave packet (24) has 
momentum cutoff at  $k \sim \Lambda^{-1}$ where $\omega_0 \Lambda \gg 1$, we
see that for our wave packet $k^2 \ll \omega_0^2$ so that we may neglect the
$-k^2$ term altogether. Hence, we again obtain Eq.(57) with $\Omega_0^2 =0, \;
\Omega^2_{\pm} = 4 \omega^2_0$ as in the plane wave case.
Therefore, the discussion following Eq. (57) applies 
also to the present case, 
including results (66-68) for constants $d_{\pm}, \; c_{\pm}$. We again find 
the radiation.
Using formula (62) and asymptotic formula (64) for the Hankel function
we obtain the following form of the radiation part of the backreaction in the 
standing wave packet case 
\begin{equation}
\delta \tilde{\chi}^{(2)} \approx (2\pi)^{1/2}  \kappa   
(4 \omega_0^2 - \kappa^2)^{- 1/4} C(\kappa)   w_2(\xi_3) 
 \sqrt{r} \cos( - 2 \omega_0\xi^0 + 2 q \xi^3
+ \sqrt{4\omega_0^2 - \kappa^2}\; r - \pi/4),
\end{equation}
It is a cylindrical wave, modulated along the vortex by the $w_2(\xi_3)$
function. The corresponding wave vector has the components
\[ (k^{\mu}) = \left( 2 \omega_0, \sqrt{4\omega_0^2 - \kappa^2}\; r^1/r,
\sqrt{4\omega_0^2 - \kappa^2}\; r^2/r, 0 \right). \]
Its 4-dimensional invariant length is equal to $\kappa^2$.

The fact that the wave front has the cylindrical shape 
is in accordance with the  conical 
character of the emitted wave in the previous example. The standing wave 
packet can  be regarded as containing $+q$ and $-q$ components with an equal 
weight -- such a superposition of conical waves (69) gives the cylindrical 
wave.

Proca wave packets moving along the vortex can be obtained by Lorentz boosts.

\section{ REMARKS} 
\paragraph{1.}Let us summarize the main points of our paper. 
In order to investigate the excited vortex fields, introduced by the Ansatz 
(2-4), we have used the perturbative expansion in the
amplitude of the excitation.  The backreaction is given as the corrections
$F^{(2)}, \; \chi^{(2)}$ to the unexcited vortex fields $F^{\rm ANO},\;
\chi^{\rm ANO}$. We have found that  $F^{(2)}$ exponentially vanishes,
but the corresponding characteristic length is large ($\sim k_0^{-1}
m_H^{-1}$),
much larger than the coherence length ($\sim m^{-1}_H$) and the penetration
depth ($\sim \kappa^{-1} m_H^{-1}$).  For the very small values of $\kappa$
assumed in our paper, there is no radiation of the scalar field from the
excited vortex. The vortex fields $F,\; \chi$ couple to 
$W_{\alpha} W^{\alpha}$ -- this composite field has an effective mass equal 
to $4\omega_0^2$. For $\kappa \ll 1/2$ this characteristc frequency 
turns out to be too small to induce radiation of the
Higgs field $F$, but it is large enough for radiation of the vector field 
$\chi$. Nevertheless, radiation of the Higgs field might appear in higher 
orders in the perturbative expansion.  The frequency of the radiative
component of $\delta \tilde{\chi}^{(2)}$, 
c.f. formulae (69), (72), is twice of that of 
the excitation field $W_{\alpha}(\xi)$.

\paragraph{2.}Perhaps the main question about our approach is whether the 
approximations 
used are correct. Rigorous convergence proofs are not known neither for the 
polynomial approximation nor for the expansion in the amplitude of 
the excitation. Confrontation of the polynomial approximation for the
unexcited  ANO vortex with a purely numerical solution shows that it
works rather well [12]. One reason is that the polynomials are
used in a finite interval of 
$r$ ($0 \leq r \leq r_0$), where the vortex fields
are expected to have a simple behaviour (e.g., one does not expect that they
oscillate like $r\sin(1/r)$ for $r \to 0$). Therefore, low order polynomials
can give quite good approximation.  

Another reason is that the scalar and vector fields in the Abelian Higgs 
model are massive ones, and therefore the vortex fields  approach the 
corresponding  vacuum values very rapidly (exponentially). Therefore, 
deviations  from the vacuum values for large $r$ can be regarded as small
corrections which one may reliably determine from linearised equations.

As for the expansion in powers of the amplitude, certainly encouraging
is the fact that the correction $F^{(2)}$ has turned out to be small for all 
values of $r$. So is $\chi^{(2)}/r$ (which is the function actually 
present in our formulae). Moreover, the exact equations (6,8,9,10) depend on 
the excitation fields $A^{\beta}, \vartheta$ in a smooth manner, 
and the corresponding
terms are just additional contributions to the other nonlinear terms which
are already present in the case of the unexcited ANO vortex. 
The presence of the excitation
fields $A_{\vartheta}, \vartheta$ in Eq. (6) does not seem to introduce any
singularity in that equation. All that suggests that
also the solution describing the excited vortex depends  smoothly on the 
amplitude of the excitation field if the amplitude is small. 

\paragraph{3.} The presence of radiation means that the excitation 
is not stable. Nevertheless,  lifetime of the small amplitude excitations 
is large because  energy of the excitation is proportional to $N^2$  
(for concreteness we are referring to the plane wave excitation) 
while the energy 
flux is of the order $N^4$. Our formulae also show that the energy flux
increases when the amplitude grows, however our approximations  
should not to be trusted when the amplitude becomes large.

\paragraph{4.} Finally, let us mention two directions for rather 
interesting  extensions of the present work. 
According to our results, the excited vortex is no longer localised within 
the penetration depth range ($(\kappa m_H)^{-1}$). This likely influences its
interaction with other vortices. The new interaction, unknown as yet, 
may turn out to have quite interesting  properties.

Also, it would be interesting to apply the polynomial approximation in order
to calculate radiation of Goldstone particles from global vortices. Such an
analytical approach would nicely complement existing numerical investigations
as well as analytical results based on Lund-Regge string model for the global
vortices presented in, e.g., [13], where one can also find references to
older papers on radiation from the global vortices.

\end{document}